\documentclass[prd,twocolumn,unsortedaddress,superscriptaddress,a4paper,nofootinbib,%
preprintnumbers,showpacs,amsmath]{revtex4-1}
\usepackage{times}
\usepackage[utf8x]{inputenc}
\usepackage{amsmath,amsbsy, amssymb}
\usepackage{enumitem}
\usepackage{dcolumn}
\usepackage{natbib}
\newcolumntype{.}{D{.}{.}{-1}}

\def\vec#1{\boldsymbol{#1}}
\usepackage[pdftex]{graphicx}
 \def\scalar#1{\raisebox{.7em}{\scalebox{1}[-1]{$\tilde{\raisebox{.7em}{\scalebox{1}[-1]{$#1$}}}$}}}
\def\scQ{\scalar{Q}}
\def\scQb{{\small \hskip 3pt $\overline{\hskip -3pt\mbox{\normalsize $\scQ$}}$}}
\def\scu{\scalar{u}}
\def\scd{\scalar{d}}

\usepackage{hyperref}
\hypersetup{
backref=true, 
pagebackref=true,
hyperindex=true, 
colorlinks=true, 
breaklinks=true, 
urlcolor= blue, 
linkcolor= blue, 
bookmarks=true, 
bookmarksopen=true, 
pdftitle={Cours L3}, 
pdfauthor={Jean-Marc Richard}, 
pdfsubject={L3} 
}

\begin{document}
\preprint{CERN-PH-TH/2013-200}
\title{Scalar Quarks at the Large Hadron Collider}
\author{Jean-Marc Richard}
\affiliation{Universit\'e de Lyon and Institut de Physique Nucl\'eaire de Lyon, IN2P3-CNRS--UCBL\\
       4 rue Enrico Fermi, 69622 Villeurbanne, France}
\email{j-m.richard@ipnl.in2p3.fr}
\author{Tai Tsun Wu}
\affiliation{Gordon McKay Laboratory, Harvard University,
Cambridge MA 02138, USA} 
 \affiliation{%
 Theory Division, CERN, CH-1211 Geneva 23, Switzerland}
\email{Tai.Tsun.Wu@cern.ch}
\date{\today}
\begin{abstract} 
    The properties of scalar quarks are studied, especially the formation
of fermionic mesons with an anti-quark.  On the basis of this
theoretical investigation
together with the experimental data, both from
last year and from this year, of the ATLAS Collaboration and the CMS
Collaboration at the Large Hadron Collider, it is 
proposed that the
standard model of Glashow, Weinberg, and Salam should 
be augmented by
scalar quarks, scalar leptons, and additional fermions.  If these
scalar quarks and scalar leptons are in one-to-one correspondence
with the ordinary quarks and ordinary leptons, either in number or in
the degrees of freedom, then there may be a fermion-boson symmetry. 
The fermion-boson symmetry obtained this way is of a different nature
from that of supersymmetry.
\end{abstract}
\maketitle
\section{Introduction}\label{se:intro}
  One year ago, in the summer of 2012, a new particle was discovered by the ATLAS
Collaboration \cite{Aad:2012tfa}  and the CMS Collaboration \cite{Chatrchyan:2012ufa} at CERN.  This particle is
almost certainly the long-sought one proposed on theoretical grounds in 1964
\cite{Englert:1964et,*Higgs:1964ia,*Guralnik:1964eu}, and is commonly referred to as the Higgs particle.

     This particle is the first observed elementary particle of spin zero.
A particle of such novel feature may open up an entirely new field of
physics. 

   In as much as the first spin-0 particle -- the Higgs particle -- has
been observed experimentally, it seems likely that this is not the only
spin-0 elementary particle and that there are other spin-0 elementary
particles.  What can these additional spin-0 particles be, and is there
any experimental indication that they may be present?

   We believe that there is such an indication.  In both \cite{Aad:2012tfa} by the ATLAS
Collaboration and \cite{Chatrchyan:2012ufa} by the CMS Collaboration, the decay rate for
\begin{equation}\label{eq:Hgg}
  H \to\gamma \gamma~, 
\end{equation}
is higher than that predicted by the standard model \cite{Glashow:1961tr,*Weinberg:1967tq,*Salam:1968rm}. It is purpose
of this paper to analyse the posssible consequences of the proposal that
this excess remains for future data.

Since this decay \eqref{eq:Hgg} proceeds through a virtual
loop, its excess is most naturally explained by the contributions of
additional charged,  heavy particles to this loop.  What are the possible properties
of these additional heavy particles?

     Since elementary particles of spin higher than 1 has never been
seen, let us consider the three cases that the additional heavy particle
for the loop is of spin 1, 1/2, and 0.

     If the spin is 1, it is likely to be a new gauge particle.  Such a
gauge particle would enlarge the gauge group of $\mathrm{U(1)\times SU(2)\times SU(3)}$ of
the standard model.  This would change the properties of the standard
model.

     If the spin is 1/2, it is likely to be a particle of the fourth
generation of quarks and leptons.  Such particles of the fourth
generation have been looked for experimentally, but none has been found so
far.

     If the two cases above of spin 1 and spin 1/2 are considered to be
unlikely, then we are left with the possibility that the additional heavy
particle is of spin 0.  This is the possibility to be investigated in this
paper. 

      Scalar quarks have been discussed in the framework of supersymmetric theories \cite{Golfand:1971iw,*Volkov:1973ix,*Wess:1974tw,*Salam:1974yz}. However, the scalar quarks under consideration here do not need to the squarks of supersymmetry. We simply assume that the scalar quarks have color $3$, and hence participate into the confining interaction of QCD. Note that scalar quarks have been considered in different contexts, for instance in some studies of chiral dynamics, confinement and diquark clustering within QCD \cite{Iida:2008cq}.

In summary, the scenario to be investigated here is the one that the
standard model is augmented by scalar quarks and scalar leptons.
\section{Some Immediate Consequences}\label{se:immediate}
Nearly a quarter century ago, one of the first results from the Large Electron-Positron collider
(LEP) at CERN was that there are three generations of quarks and
leptons \cite{ALEPH:2005ab}.  
This implies that the $Z$ cannot decay into a pair of scalar
quarks or scalar leptons, meaning that the mass of each scalar quark and
each scalar lepton must be larger than half of the $Z$ mass, i.e., 46\;GeV$/c^2$ (Strictly speaking, this limit should be slightly reduced because of phase space). 

On the other hand, if the scalar quarks and the scalar leptons are
responsible for the excess in the decay process \eqref{eq:Hgg}, then they cannot be
too heavy, or more precisely cannot be much heavier than the $Z$.  If only
these scalar quarks and the scalar leptons are added to the standard
model, then both the lightest scalar quark and the lightest scalar lepton
are stable.  One of the important questions to be answered in the present
paper is: 
Is it possible that the LHC experiments have failed to  detect  such a stable lightest scalar quark with a mass of the order of magnitude of that of the $Z$?

     The answer is No. In particular, the CMS collaboration has published a thorough search for new charged particles and set stringent constraints \cite{Chatrchyan:2013oca}. We will show below that if there is a long-lived scalar quark of color $3$, with either electric charge $2/3$ or $-1/3$, there should be at least a long-lived new hadron that is charged and thus should have been seen, contradicting the CMS result. 
Therefore, there must be some additional new
particles for the lightest scalar quark to decay into.  At the moment,
there does not seem to be enough information from the data in \cite{Aad:2012tfa} and \cite{Chatrchyan:2012ufa}
to pin down what these additional new particles may be.

     Scalar quarks can of course be pair produced.  Since scalar quarks
are confined just like ordinary quarks, a scalar quark, denoted $\scQ$, thus produced picks
up an anti-quark to form a new kind of meson.  This anti-quark can be
either an ordinary anti-quark  of spin 1/2, especially the  $u$ or the $d$
anti-quarks, or a scalar anti-quark \scQb\ of spin 0.  Since a scalar anti-quark
has a mass of at least half of that of $Z$, it is much heavier than that of
the $u$ or $d$ anti-quarks.  While both possibilities exist, it is more likely
that the first detected bound state will consist of a scalar quark together with an $u$ or $d$
anti-quark, say $(\scQ\bar q)$, and that the ``scalaronium'' ($\scQ$\scQb) 
will be more difficult to be produced.  In the case of a $(\scQ\bar q)$ bound state, the resulting meson is different from the known
mesons in that it is a fermion. Similarly the first baryon including a scalar quark is very likely of the type $(\scQ qq')$ with two light ordinary quarks. It is a boson. 
%
%
%
\section{Hadrons containing a scalar quark}
Let us consider some hadrons formed by a scalar quark $\scQ$ with ordinary
quarks $q$ ($u$, $d$, $s$, $c$ or $b$)  or antiquarks $\bar q$.

The first set of new hadrons contains a single scalar quark. For any given $q$, the ground state of $(\scQ\bar q)$ is not degenerate, and has an angular momentum $j=1/2$. The same holds for its radial excitations.  The orbital excitations with angular momentum $\ell>0$ are split by spin-orbit forces into a state with $j=\ell-1/2$ and another one with $j=\ell+1/2$.

For any given $q$, the lightest $(\scQ qq)$ baryon has spin $j=1$, and is slightly pushed up by the repulsive chromomagnetic interaction among the two $q$ quarks. If one introduces two ordinary quarks $q$ and  $q'$ with $q\neq q'$, the lightest $(\scQ q q')$ baryon has spin 0, and  above lies a spin~1 baryon. The spacing can be estimated from the phenomenology of hyperfine interactions in baryons \cite{DeRujula:1975ge,LeYaouanc:1976kp,*Klempt:2009pi}.  As discussed in some detail in Sec.~\ref{se:baryons}, one can in particular estimate \footnote{Hereafter, the name of particle also denotes its mass. For example, here $(\scu ud)_{S=1}$ means the mass of the particle that consists of the scalar
quark $\scu$, together with the ordinary quarks u and d, in the spin $S=1$ state.}
$(\scd ud)_{S=1}-(\scd ud)_{S=0}\simeq 200\;\mathrm{MeV}$
for the lightest baryons. This means that the isovector, vector state will easily decay into the isoscalar, scalar one by pion emission.

In the second set come hadron states containing two scalar quarks. The scalaronium  ($\scQ
$\scQb) certainly deserve special discussions. The spectrum of scalaronium has been discussed in several papers \cite{Nanopoulos:1983ys,*Herrero:1987xj,*Barger:2011jt}, by extrapolation of the models describing the charmonium and bottomonium states. 
These papers made explicit reference to supersymmetry. 

In this second set, there are also baryons with two scalar quarks. If $\scQ\neq \scQ'$, the ground state of the $(\scQ\scQ' q)$ has spin 1/2  and is not degenerate. It can be studied in the Born--Oppenheimer limit, as done for double-charm baryons \cite{Fleck:1989mb}, in which the effect of the light quark and gluon field are integrated out and generate an effective $\scQ\scQ'$ interaction. For the baryons with two identical scalar quarks, $(\scQ\scQ q)$,  the   $(\scQ\scQ)$ pair is in a $\bar 3$ state, which is an antisymmetric coupling of two colour $3$ objects, and thus the orbital momentum among the two heavy scalar quarks is $\ell=1$. Hence the lightest $(\scQ\scQ q)$ have $j=1/2$ and $j=3/2$, split by spin-orbit forces.

The second set also includes $(\scQ\scQ\bar q\bar q)$ tetraquarks, whose mass is lighter than the threshold made of two $(\scQ\bar q)$ mesons, thanks to the strong attraction between the two heavy scalar quarks, which are in a relative P-wave. This is the same mechanism that binds ordinary tetraquarks $(QQ\bar q\bar q)$ in the limit of a large quark-to-antiquark mass ratio, see, e.g.,~\cite{Vijande:2009xx,*Karliner:2013dqa} and refs.\ therein.

In the far future, one should aim at detecting baryons with three scalar quarks. 
If the scalar quarks are all different, then the ground state of $(\scQ\scQ'\scQ'')$ is a non-degenerate scalar, with any possible radial and orbital excitations. If $\scQ\neq\scQ'$, the ground state of $(\scQ\scQ\scQ')$ is a non degenerate $j=1$ state, due to the Bose statistics of the two $\scQ$. If the masses are such that $\scQ<\scQ'$, the first excitation is likely an $\ell'=1$ promotion of the relative motion of $\scQ'$ with respect to the two $\scQ$, and this will give $j=0,\,1$ or $2$, with some spin-orbit splitting among them. If $\scQ>\scQ'$, then the first excitation is in the relative motion between the two $\scQ$.

Let $\scQ$ be the lightest of the scalar quarks, then the first baryon with three scalar quarks is of the type $(\scQ\scQ\scQ)$ with an antisymmetric color wavefunction, and thus an antisymmetric orbital wave function, with $\ell=1$ among the first two constituents and also $\ell'=1$ for the relative motion of the third one, and an overall angular momentum $j=1$. In the particular case of harmonic confinement, the wave function is of the type $\vec\rho\times\vec\lambda\,\exp[-\alpha(\vec\rho^2+\vec\lambda^2)]$, where $\vec\rho=\vec r_2-\vec r_1$ and $\vec\lambda=(2\,\vec r_3-\vec r_1-\vec r_2)/\sqrt3$ are the usual Jacobi variables. In the ordinary quark-model, this orbital wave function is associated to the long sought $[20,1^+]$ representation in the specific jargon of this field, a state just with two levels of excitation above the nucleon-$\Delta$ multiplet.   See, for instance, the reviews \cite{Hey:1982aj,*Richard:1992uk}. The lack of experimental evidence for this state is one of the motivations for the quark-diquark model of ordinary baryons~\cite{Anselmino:1992vg}. It is amazing that the antisymmetric orbital wavefunction of three confined constituents reappears in the context of scalar quarks.
\section{Lightest meson  with a scalar quark}\label{se:lightest-meson}
For the experimental search of scalar quarks, it is important  to know whether the lightest meson containing  a 
scalar quark is neutral or charged. 
Let $\scd$ denote a scalar quark $\scQ$  with electric charge $-1/3$, and $\scu$ one with
electric charge $+2/3$.  It is important to know whether
 the lightest meson is the neutral  $(\scd\bar d)$ or the charged $(\scd\bar u)$, and similarly the charged $(\scu\bar d)$  or the neutral $(\scu\bar u)$ meson. 
In this section, we calculate the masses of $(\scu \bar q)$ and $(\scd \bar q)$ mesons assuming that the scalar quarks $\scu$ and $\scd$ are stable or long-lived, and analyze whether this results into stable or long-lived charged mesons that could be seen in the LHC detectors.

We therefore proceed to analyze to which extent our present understanding of the isospin-breaking mass differences allows us an extrapolation toward the mesons containing a scalar quark. In particular, the mass differences $D^+-D^0$  and $B^0-B^-$ are now  well measured \cite{Beringer:1900zz,Aaij:2013uaa}, with the results
\begin{equation}\label{eq:iso-D-B}
\begin{aligned}
D^+-D^0&=4.76\pm0.06\,{\rm MeV}~,\\
B^0-B^-&=0.32\pm0.06\,{\rm MeV}~. 
\end{aligned}
\end{equation}

There is a long history of studies of the mass splittings among members of the same hadron multiplets. Some milestones references are given in \cite{Godfrey:1985sp}. For our purpose, it is important to note that the splittings \eqref{eq:iso-D-B} can be well reproduced by quark models that include a relativistic kinematics, a flavor-independent central potential, and a spin-spin interaction that depends explicitly upon the constituent masses.
In such quark models, the mass difference $(Q\bar d)-(Q\bar u)$ between the isospin partners originates from
\begin{enumerate}[label={\sl \alph*)}]\itemsep-2pt
 \item \label{iso1} the change $m_d-m_u$ in the constituent masses,
\item\label{iso2} the induced change of the binding energy, i.e., the change in the expectation value of the kinetic energy and central potential,
\item\label{iso3} the change in the spin--spin term,
\item\label{iso4} the electrostatic energy (the magnetic term is very small).
\end{enumerate}
 There are cancellations, making the adjustment of the parameters delicate. In particular, the effects \ref{iso1} and \ref{iso2} are opposite. In non-relativistic models, it might even happen that \ref{iso2} overcome \ref{iso1}. This is not the case with a relativistic form of kinetic energy such as $\sqrt{\vec p^2+m^2}$.
For the $D$, the electrostatic contribution \ref{iso4} supplements the mass effect \ref{iso1} and \ref{iso2}, but for the $B$, there is a cancellation. This explains the hierarchy observed in Eq.~\eqref{eq:iso-D-B}. 

For $D$ and $B$ mesons, the spin--spin effect \ref{iso3} plays a minor role. First, there is  an overall $1/m_c$ or $1/m_b$ factor  in most models. Also, when the light quark mass increases, the strength of the hyperfine interaction decreases, but the wave function becomes more compact, and this increases the expectation value of the contact or contact-like terms associated to the hyperfine interaction. As noted, e.g., in \cite{Goity:1992kk}, it is observed experimentally that $D_s^*-D_s\simeq D^*-D$ and $B_s^*-B_s\simeq B^*-B$, a kind a SU(3) restoration. Thus, if the hyperfine interaction does not change significantly when $u$ is replaced by $s$, it will remain almost constant when $u$ is replaced by $d$. 

As a consequence, if the heavy quark is made scalar, the removal of the hyperfine term will not affect significantly the isospin splitting. Moreover, with $c$ and $b$, we are already in a regime of heavy-quark symmetry for the wave function. 
In a non-relativistic language, one can say that the reduced mass is nearly saturated by the light-quark mass. This means that when the scalar quark is made heavier, say from a few GeV to 46\,GeV or more, the Coulomb energy and  the binding energy do not change much.
On this basis, one expects that all quark models will give
\begin{equation}
 \label{eq:rough-prediction}
\begin{aligned}
(\scu\bar u)-(\scu\bar d)&\simeq (c\bar u)-(c\bar d)~,
\\
(\scd\bar u)-(\scd\bar d)&\simeq (b\bar u)-(b\bar d)~,
\end{aligned}
\end{equation}
where the approximate equalities \eqref{eq:rough-prediction} mean that
\begin{equation}
 \label{eq:detailed-prediction}
\begin{aligned}
\left|\left[(\scu\bar u)-(\scu\bar d)\right]-\left[ (c\bar u)-(c\bar d)\right]\right|&\lesssim 1\,\mathrm{MeV}~,\\
\left|\left[(\scd\bar u)-(\scd\bar d)\right]-\left[ (b\bar u)-(b\bar d)\right]\right|&\lesssim 1\,\mathrm{MeV}~.
\end{aligned}
\end{equation}
Indeed, these inequalities \eqref{eq:detailed-prediction} are verified in the explicit computation presented below.  

In order to make some quantitative predictions, we adopt an explicit  
 semi-relativistic model, similar to the ones used, e.g., in \cite{Stanley:1980zm,*Carlson:1982xi,*Basdevant:1984rk,Godfrey:1985sp}, 
namely
\begin{equation}
 \label{eq:H}
\begin{aligned}
&H=H_0+V_{ss}~,\qquad 
V_{ss}=d\,\frac{\vec\sigma_1.\vec\sigma_2}{m_1\,m_2}\,\tilde\delta(\vec r)~,\\
&H_0=\sqrt{\vec p^2+m_1^2}+\sqrt{\vec p^2+m_2^2}-a/r+b\,r+c~,
\end{aligned}
\end{equation}
where $a$, $b$, and $c$ are independent of the quark masses $m_i$, as suggested by flavor independence. The term $V_{ss}$ is  a smeared version of the contact interaction which describes the hyperfine splittings \cite{DeRujula:1975ge}, analogous to the Breit--Fermi interaction in QED. In practice, we use
\begin{equation}
 \label{eq:tilde-delta}
\tilde\delta(\vec r)=(\mu/\pi)^{3/2}\,\exp(-\mu\,r^2)~,\quad
\mu=\genfrac{(}{)}{}{}{2\,m_1\,m_2}{m_1+m_2}^f~,
\end{equation}
where  $f$ is an empirical parameter close to 2, so that the range  of the smearing is of the order of magnitude of the de~Broglie wavelength of the relative motion. Many refinements can be envisaged, as, e.g., in \cite{Godfrey:1985xj}, but we start from the simplest version of the model and discuss some possible corrections later in this section. 

It is obviously impossible to fit all meson levels with any flavor content with such a simple model, so we focus on reproducing the lightest quarkonium levels, and the ground states in the heavy-light sector.
To solve the Schr\"odinger equation $H\,\Psi=M\,\Psi$, we expand the wave function into Gaussians, so that the matrix elements can be computed either in position or in momentum space. Explicitly, 
\begin{equation}
 \label{eq:gauss} \Psi=\sum_{i=1}^N\gamma_i\,\exp(-\alpha_i\,r^2/2)~,
\end{equation}
with the non-linear parameters $\alpha_i$ imposed to follow an arithmetic progression, to avoid ambiguities in the minimization, following the strategy of  \cite{Hiyama:2003cu}. For a variant, see \cite{Suzuki:1998bn}. The results are presented in Tables \ref{tab:para}, \ref{tab:resu1}  and \ref{tab:resu2}.
\begin{table}[htp]
 \centering
 \caption{\label{tab:para} Parameters of the quark model of Eqs.~\eqref{eq:H} and \eqref{eq:tilde-delta}. The units are powers of GeV}
\begin{ruledtabular}
\begin{tabular}{ccccccc}
\multicolumn{1}{c}{$m_d$} & \multicolumn{1}{c}{$m_u$} & \multicolumn{1}{c}{$m_c$} & \multicolumn{1}{c}{$m_b$} &
 \multicolumn{1}{c}{$m(\scQ)$} & \multicolumn{1}{c}{$m(\scQ')$} \\
  0.305    &  0.300     &    1.500   &   4.900 &    100. & 200. \\
\end{tabular}
\vskip 5pt
\begin{tabular}{ccccc}
  $a$ &  $b$  &   $c$  &  $d$ &  $f$   \\
0.3   &  0.2  &  $-0.38$ & 1.39 &  1.5   \\
\end{tabular}
\end{ruledtabular}
\end{table}
\begin{table}[htp]
 \centering
 \caption{\label{tab:resu1}Contributions (in MeV) to  the isospin-breaking mass difference of $D$ and $B$ mesons in a specific version of the model of Eqs.~\eqref{eq:H} and \eqref{eq:tilde-delta}. $h_0=H_0-m_1-m_2$ is the sum of kinetic energies and central potential. } 
\begin{ruledtabular}
\begin{tabular}{c.....c}
& \multicolumn{1}{c}{$m_d-m_u$} & \multicolumn{1}{c}{$\langle  h_0 \rangle$} &  \multicolumn{1}{c}{$\langle V_{ss}\rangle$} & 
\multicolumn{1}{c}{Coulomb} & \multicolumn{1}{c}{Total} & \multicolumn{1}{c}{Exp.\ \protect\cite{Beringer:1900zz}}\\
$D_d-D_u$ &  5. & -2.5 & -0.08 &  3.5  & 5.9  &$\phantom{-}4.76\pm0.10$ \\
$B_d-B_u$ & 5. &  -2.3 & -0.5 & -1.9 & 0.2 & $-0.32\pm0.06$\\
\end{tabular}
\end{ruledtabular}
\end{table}

\begin{table}[htp]
 \centering
 \caption{\label{tab:resu2}Isospin breaking mass differences in the model (in MeV), for two values of the mass of the scalar quark (In GeV)}
{
\begin{ruledtabular}
\begin{tabular}{ccc}
$m(\scQ)$ & $(\scu\bar d)-(\scu\bar u)$ & $(\scd\bar d)-(\scd\bar u)$ \\
100.&     6.6  &    $0.06$   \\
200.&     6.6 &    $0.04$   \\
\end{tabular}
\end{ruledtabular}
} 
\end{table}

\begin{figure}[ht!]
 \centering
 \includegraphics[width=.90\columnwidth]{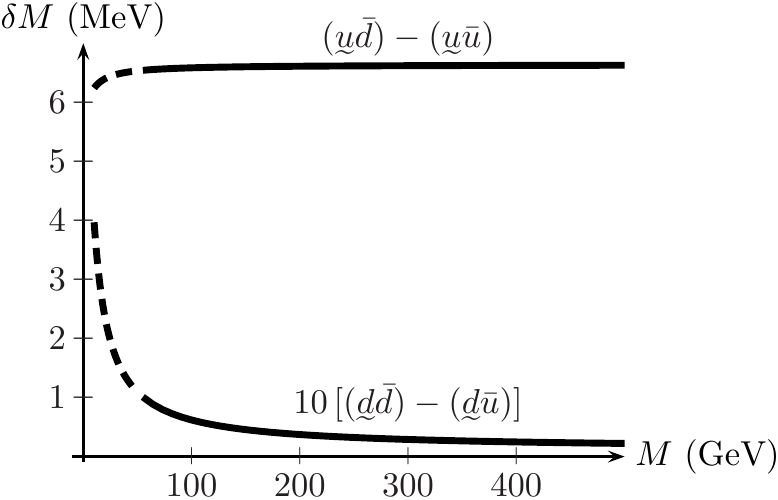}
 \caption{\label{fig:M} Mass difference $\delta M=(\protect\scu\bar d)-(\protect\scu\bar u)$ and $\delta M=(\protect\scd\bar d)-(\protect\scd\bar u)$ (multiplied by 10) in our model, as a function of the scalar-quark mass $M$. The dotted part corresponds to low values of $M$ that has been experimentally excluded.%
}
\end{figure}

Comments are in order. 
\begin{itemize}
\item The variational calculation  converges rapidly. The results in the Tabs.\ \ref{tab:resu1} and \ref{tab:resu2} correspond to $N=6$ Gaussians. The  change with respect to $N=5$ is only $0.4\%$ for $D^+-D^0$ and $-3\%$ for $B^0-B^-$.
 \item 
The results are very stable. One can play with, e.g., the difference of constituent masses, $m_d-m_u$, or the parameters entering the potential, but it always turns out that $D^+-D^0$ is about 5\,MeV, while $|B^0-B^-|$ is less than 1\,MeV.
For instance, if we repeat our calculation with $m_d=310\,$ MeV instead of $305\,$MeV, the results are changed to  (in MeV)
$(c\bar d)-(c\bar u)=7.3$, $(\scu\bar d)-(\scu\bar u)=8.8$, $(\scu\bar d)-(\scu\bar u)=8.9\,$MeV (for 100\,GeV or 200\,GeV scalar quark, respectively), and $(b\bar d)-(b\bar u)=2.6$, $(\scd\bar d)-(\scd\bar u)=2.3$, $(\scd\bar d)-(\scd\bar u)=2.3\,$MeV. The fit to $D$ and $B$ mesons somewhat deteriorates, but the validity of \eqref{eq:rough-prediction} remains. 
\item 
If one introduces a form factor for the Coulomb interaction, as per Eq.\ (4) of Godfrey and Isgur \cite{Godfrey:1985sp}, the electrostatic energy is considerably reduced, by a factor of about 2,
and it becomes difficult to get a good agreement  for both $D^+-D^0$ and $B^0-B^-$. In our case, such a form factor would give $D^+-D^0= 2.9\,$MeV instead of 4.9, and $B^0-B^-=1.5\,$MeV instead of 0.4. However, the isospin splittings remain of the same order of magnitude, and the inequalities \eqref{eq:rough-prediction} and \eqref{eq:detailed-prediction} remain valid for the extrapolation to scalar quarks. One motivation in \cite{Godfrey:1985sp} is to account for ``relativistic nonlocalities'', but this effect is in principle an output rather than an input of the relativistic equations such as \eqref{eq:H}.
\item The mass differences evolve very smoothly with the mass of the scalar quark $\scQ$. This is illustrated in Fig.~\ref{fig:M}.
\item
If the scalar quark $\scu$ is long lived, 
the lightest charged meson, $(\scu\bar d)$,  could decay weakly into  $(\scu\bar u)$. But with an energy release of about 6\,MeV, the lifetime for such a decay would be very long, so both $(\scu\bar d)$ and  $(\scu\bar u)$ would be long lived. 
\item
Similarly, if the scalar quark $\scd$ is long lived, 
the beta decay from  $(\scd\bar d)$ to  $(\scd\bar )$, or \textsl{vice-versa}, is either kinematically forbidden or  very slow. So both $(\scd\bar d)$ to  $(\scd\bar )$ would be long lived. 
\item
Therefore, as no charged track has been seen in the CMS detector \cite{Chatrchyan:2013oca}, one can exclude a scenario where the standard model is augmented by scalar quarks and scalar letpons only.
\end{itemize}

\section{Lightest baryon  with a scalar quark}\label{se:baryons}
Here the situation is simpler than in the meson case. For each scalar quark $\scQ=\scu$ or $\scd$, the lightest baryons are $(\scQ uu)$, $(\scQ dd)$ and $(\scQ ud)$. For the two former baryons, the ground state, with mainly an overall s-wave in the relative motion, correspond to the identical light quarks being in a color $\bar 3$, spin 1 and isospin 1, state. For the latter, there is one state, say $\Sigma[\scQ]$, with $(u, d)$ in a spin 1 and isospin 1 state, and another one, $\Lambda[\scQ]$, with spin 0 and isospin 0. We expect the vector, isovector states $(\scQ uu)$, $(\scQ dd)$ and $\Sigma[\scQ]=(\scQ ud)$ to be very close, and about 200\,MeV above the scalar, isoscalar $\Lambda[\scQ Q]=(\scQ ud)$, because the $\Sigma-\Lambda$ mass difference comes from the light-quark part of the baryons, and thus should be very similar for $\Sigma_c-\Lambda_c$. More precisely, if the spin-spin interaction in $(Qqq)$ is treated at first order, one gets, in an obvious notation
\begin{equation}
 \label{eq:Lambda-Sigma-1}
\begin{aligned}
\Lambda_Q&=M_0-3\,\delta_{qq}~,\\
\Sigma_Q&=M_0+\delta_{qq}-4\,\delta_{Qq}\,\\ 
\Sigma_Q^*&=M_0+\delta_{qq}+2\,\delta_{Qq}~,
\end{aligned}
\end{equation}
and thus a fictitious charmed or bottom baryon in which the spin-spin interaction of the heavy quark is switched off would correspond to
\begin{equation}
 \label{eq:Lambda-Sigma-2}
\langle\Sigma_Q\rangle=\frac{2\,\Sigma_Q^*+\Sigma_Q}{3}=M_0+\delta_{qq}~,
\end{equation}
and from the know masses of charmed and beauty baryons~\cite{Beringer:1900zz}, one obtains
\begin{equation}
 \label{eq:Lambda-Sigma-3}
\langle\Sigma_c\rangle-\Lambda_c\simeq 212\,\mathrm{MeV}~,\quad \langle\Sigma_b\rangle-\Lambda_b\simeq 207\,\mathrm{MeV}~.
\end{equation}
This spacing is remarkably stable when going from charm to beauty. It is thus very easy to extrapolate to 
 \begin{equation}
 \label{eq:Lambda-Sigma-4}
(\scQ qq)_{I=1,J=1}-(\scQ qq)_{I=0,J=0}\sim 200\,\mathrm{MeV}~.
\end{equation}

In short, the chromomagnetic mechanism by which $\Lambda_Q$ is lighter than $\Sigma_Q$ for the quarks $Q=s$, $c$ and $b$, implies that the lightest baryons containing a scalar quark are the charged, scalar, isoscalar $(\scu ud)$ and the neutral, scalar, isoscalar 
$(\scd  ud)$.
\section{Further properties of scalar quarks}
The results obtained so far about the scalar quarks are largely
independent of their detailed properties.  What has been used consists
mainly of their existence, their being color~$3$, and their masses being at
least half of the mass of $Z$.

In this Section, some further possible properties of the scalar quarks are
to be discussed briefly.

(A)\ 
Since these scalar quarks are unstable, how do they decay?  There are
two distinct types of decays.

\quad{\sl i)}\ Similar to ordinary quarks, the scalar quarks have the decay mode
\begin{equation}\label{eq:decay1}
 \text{heavier scalar quark}\to \text{lighter scalar quark} + W~,
\end{equation}
where the $W$ may be real or virtual.  Of course, the heavier and lighter
quarks must have different charges.

\quad{\sl ii)}\ Since the lightest scalar quark is not stable, there is also the decay
mode
\begin{equation}\label{eq:decay2}
             \text{scalar quark} \to \text{quark} + \text{fermion}~,
\end{equation}
where the fermion on the decay product is a new particle that has not been
observed experimentally,  see Sec.~\ref{se:immediate}.  This fermion can of course be real
or virtual.  In MSSM, this fermion can be higgsino, wino, or zeno. 

(B)\  How many different scalar quarks are there?  There is no way at
present to answer this important question, but there are two natural
guesses.

\quad {\sl i)}\  There are six scalar quarks and six scalar leptons;

\quad {\sl ii)}\  There are twelve scalar quarks and twelve scalar leptons.

In {\sl i)} there is one scalar quark or lepton for each fermion in the standard
model; in {\sl ii)} the number of degrees of freedom for the scalar quarks and
leptons is the same as that for the ordinary quarks and leptons.  In MSSM,
{\sl ii)} holds but not {\sl i)}.

(C)\   The starting point of the present consideration is the branching
ratio for the decay \eqref{eq:Hgg}; it requires the masses of several scalar
quarks and/or charged scalar leptons to be of the order of the $Z$ mass.  In the
case of quarks, the qualitative features of the CKM matrix has been
attributed to the large differences of masses for the three generations of
quarks \cite{Lehmann:1995br}.  If this argument holds also for scalar quarks, then it may be
expected that several off-diagonal elements of the scalar-quark CKM matrix
elements are large in the sense of being of the order of~1.  This is
likely to lead to interesting physics for the scalar quarks.

(D)\  Quarks are seen experimentally as jets.  How do the scalar-quark jets
differ from quark jets?  It is believed that a major difference is due
simply to the fact the masses of the scalar quarks being much large than
those of the first two generations of quarks.

Let us compare a scalar-quark jet with, say, a $b$ quark jet.  In both
cases, pairs of light quarks are produced from the sea.  Thus the leading
$b$ quark becomes a $B$ meson, and the leading scalar quark $\scQ$ becomes a $(\scQ \bar{q})$ meson.  
Both the $B$ meson and the $(\scQ \bar{q})$ meson decay.  Since the
mass of the $B$ meson is only about 5\,GeV, its decay products and those of
the mesons from the sea have momenta in nearly the same direction; this is
how a $b$ jet is formed and looked for.  On the contrary, since the $(\scQ \bar{q})$
meson is much heavier, its decay products have in general momenta in
different directions.  Therefore, a scalar-quark jet has in general much
more complicated structure and cannot be found by the usual jet finding
programs.  In fact, it is perhaps more accurately described as consisting
of several jets, not a single jet.

    (E)\ Let the above considerations be combined with the very recent
results from the ATLAS and CMS Collaborations \cite{Aad:2013wqa,*MingshuiChenfortheCMS:2013fba,*CMS2013} together with those from the
CDF and D0 Collaborations \cite{Aaltonen:2012qt}.  These experimental results indicate
that the couplings of the Higgs particle \cite{Aad:2012tfa,Chatrchyan:2012ufa} to the $Z$, the  $W$, the
top quark and the bottom quark are all in agreement with those of the
standard model \cite{Glashow:1961tr,*Weinberg:1967tq,*Salam:1968rm}.  If this agreement is verified to higher accuracies in the future, there are important implications such as
follows.

    The first implication is that there is very little room for a second
Higgs particle to couple also to these $Z$, $W$, $t$, and $b$.  It may be
recalled that, in super-symmetric theories such as the MSSM, there are
necessarily two Higgs doublets, where, in the simplest version, one
couples to the top quark while the other couples to the bottom quark.

    Consider this first implication together with the (B) above, where there
is a pairing between quarks and scalar quarks, and also leptons with
scalar leptons through either {\sl i)} or {\sl ii)}.  Such pairings mean that there is
some form of fermion-boson symmetry, but this symmetry may well be of a
different nature than that of super-symmetry, which is based on graded Lie
algebra.

    Let us speculate what this fermion-boson symmetry may be.  From the
(C) above, the masses of the scalar quarks and scalar leptons are
likely to be much higher than those of the quarks and leptons, except that
of the top quark.  In other words, the structure of the masses for the
scalar quarks and leptons are certainly qualitatively different from that
of the ordinary quarks and leptons.  It would be most interesting to learn
what the mass relations are and where they may come from, but this is
beyond reach at present.
    On the experimental level, the main issue is of course to figure out
ways to observe these scalar quarks and scalar leptons.  A step in
that direction may be searching for scalar-quark jets as distinct from the
jets from ordinary quarks, as discussed in (D) above.  Searching for
scalar leptons may be equally important or perhaps even more important,
and one way to carry out this search is through the decays
\begin{equation}\label{eq:slepton-decay}
           \text{heavier scalar lepton} \to \text{lighter scalar lepton} + W~,
\end{equation}
where the $W$ may be either real or virtual, a process very similar to that
of ordinary leptons.

    On a most fundamental level, it is the purpose of the present work to
think about, on the basis of the experimental results from the Large
Hadron Collider and the Tevatron Collider, what this fermion-boson
symmetry might be.  In order to have any such symmetry, it is of course
necessary to have three generations of scalar quarks and scalar leptons. 
However, it is not sufficient merely to augment the standard model with
the scalar quarks and the scalar leptons, and one, or more likely several,
new fermions are needed -- See (A) above.  There is at present essentially
no information about these new fermions, and an interesting exercise is to
list the various attractive possibilities for them and then compare with
future experimental results.

\section{Conclusion and outlook}\label{se:concl}
     Last year, there were two epoch-making papers, one by the ATLAS
Collaboration  \cite{Aad:2012tfa}  and one by the CMS Collaboration \cite{Chatrchyan:2012ufa}, describing the
discovery of the Higgs particle \cite{Englert:1964et,*Higgs:1964ia,*Guralnik:1964eu}.  It is the purpose of the present
paper to discuss some of the possible consequences of the data presented
in these two papers.

     The starting point is the proposal that, with the discovery of the
first spin-0 elementary particle, there may well be other spin-0
elementary particles.  These additional spin-0 particles can be used to
explain the observed excess on the decay of the Higgs particle
into two photons \cite{Aad:2012tfa,Chatrchyan:2012ufa}.  Later data from the ATLAS Collaboration \cite{Aad:2013wqa,*MingshuiChenfortheCMS:2013fba,*CMS2013}
and from the CDF and D0 Collaborations \cite{Aaltonen:2012qt} are then used to reach the
conclusion that these spin-0 particles are not additional Higgs particles,
but rather scalar quarks and scalar leptons, which remain to be observed
experimentally.

     If the standard model \cite{Glashow:1961tr} is augmented only by these scalar quarks
and scalar leptons, then the lightest scalar quark is necessarily stable. 
It is shown in Sec.~\ref{se:lightest-meson} that this stable lightest scalar quark is
incompatible with the data from the CMS Collaboration \cite{Chatrchyan:2013oca}.  This means
that, in addition to the scalar quarks and the scalar leptons, there must
be additional fermions which also remain to be observed experimentally.

     At present there is no way to answer the important question of how
many scalar quarks and scalar leptons there are.  However, the
natural guess is that these scalar particles are in one-to-one
correspondence with the known fermions in the standard model, either in
number or in the degrees of freedom.  If this is the case, then it
suggests a fermion-boson symmetry.

     Since there is no additional Higgs particle, this fermion-boson
symmetry is necessarily of a fundamentally different nature from that of
supersymmetry, including MSSM in particular.  It is thus seen that the
discovery of the Higgs particle last year \cite{Aad:2012tfa,Chatrchyan:2012ufa} not only completes the
list of particles in the standard model, but also shows a possible
direction to go beyond the standard model.  In other words, it is likely
that this discovery of the Higgs particle will open up a new and exciting
era of particle physics \cite{riordan2012higgs,edition2013extreme}.

\vspace*{0.4cm}
\acknowledgments
We are very grateful to Andr\'e Martin  and Jacques Soffer for several fruitful and enjoyable discussions that motivated this investigation.
JMR would like to thank Suzanne Gascon,  Bernard Ille and Louis Sgandurra for very useful information. The work of JMR is partially supported by the Labex-LIO (Lyon Institute of Origins). 
%
%
\end{document}